\documentclass[10pt,twocolumn,english,prd,superscriptaddress,nofootinbib,preprintnumbers,showpacs,floatfix]{revtex4-2}

\usepackage{xcolor}
\usepackage[utf8]{inputenc}
\usepackage{amsmath}
\usepackage{amsfonts}
\usepackage{amssymb}
\usepackage{graphicx} 
\usepackage[caption=false]{subfig}
\graphicspath{ {figures/} }
\usepackage[usenames,dvipsnames]{xcolor}
\usepackage{hyperref}   
\hypersetup{
    colorlinks=true, 
    citecolor=MidnightBlue,
    linkcolor=MidnightBlue,
    urlcolor=Cyan}
\usepackage{tikz}
\usepackage{verbatim}
\usepackage{soul}

\definecolor{purple}{rgb}{1,0,1}
\definecolor{lime}{HTML}{A6CE39}

\begin{document}

\title{Scale Invariance, Variety and Central Configurations}

\author{Maria I. R. Lourenço}
\email{fc56407@alunos.fc.ul.pt}
\affiliation{Instituto de Astrof\'{i}sica e Ci\^{e}ncias do Espa\c{c}o, Faculdade de Ci\^{e}ncias da Universidade de Lisboa, Edifício C8, Campo Grande, P-1749-016 Lisbon, Portugal}

\author{Julian Barbour} 
\email{julian.barbour@physics.ox.ac.uk}
\affiliation{College Farm, The Town, South Newington, Banbury, OX15 4JG, UK}

\author{Francisco S. N. Lobo} 
\email{fslobo@ciencias.ulisboa.pt}
\affiliation{Instituto de Astrof\'{i}sica e Ci\^{e}ncias do Espa\c{c}o, Faculdade de Ci\^{e}ncias da Universidade de Lisboa, Edifício C8, Campo Grande, P-1749-016 Lisbon, Portugal}
\affiliation{Departamento de F\'{i}sica, Faculdade de Ci\^{e}ncias da Universidade de Lisboa, Edif\'{i}cio C8, Campo Grande, P-1749-016 Lisbon, Portugal}

\date{\LaTeX-ed \today}

\begin{abstract}

Scale invariance has received very little attention in physics. Nevertheless, it provides a natural conceptual foundation for a relational understanding of the universe, where absolute size loses meaning and only dimensionless ratios retain physical significance. We formalize this idea through the $N$-body problem, introducing a scale-invariant function---the variety, $V$---built from the square root of the center-of-mass moment of inertia and the Newtonian potential. Critical points of $V$, known as central configurations, correspond to special particle arrangements that preserve their shape under homothetic collapse or expansion. Numerical exploration of these critical points reveals that even slight deviations from the absolute minimum of $V$, which corresponds to a remarkably uniform configuration, lead to the spontaneous formation of filaments, loops, voids and other patterns reminiscent of the cosmic web. This behavior is a consequence of the intrinsic structure of shape space---the space of configurations modulo translations, rotations and dilatations---in which regions of higher variety act as attractors. Our results suggest that scale-invariant dynamics not only captures the relational nature of physical laws but also naturally generates organized patterns, offering a novel perspective on the formation of cosmic structures and on the emergence of a gravitational arrow of time from scale-invariant, relational dynamics.
\end{abstract}

\maketitle

\def\HMS{{\scriptscriptstyle{\rm HMS}}}

{\it Introduction}:
Symmetry principles have long guided the formulation and interpretation of physical laws. Among them, \emph{scale invariance} occupies a peculiar status: although central in critical phenomena, renormalization group theory and certain field-theoretic contexts, it has seldom been treated as a foundational principle in classical dynamics or cosmology, appearing instead as an accidental or approximate feature rather than an ontological commitment.

Recent work has begun to explore the consequences of treating scale invariance as a fundamental principle. In particular, reformulating Newtonian gravity in scale-invariant variables reveals a rich and previously hidden structure in its solution space \cite{barbour2014_arrow}. Eliminating absolute space, time and scale leads naturally to dynamics on shape space where an $N$-body system evolves only through shapes. Within this framework, small initial non-uniformities generically drive the system toward increasingly complex configurations, exhibiting clustering, filamentary networks and hierarchical structures. 

In this work we normalize the Newtonian potential by the square root of the center-of-mass moment of inertia, defining the resulting function as the \emph{variety}, a scale-invariant quantity that governs the system’s intrinsic dynamics. The critical points of this function---both minima and saddle points---are well known in $N$-body dynamics and referred to as \emph{central configurations}. By studying the properties of the variety and its critical points, we show that scale invariance and relational dynamics naturally give rise to structure formation. These structures emerge without additional forces, stochastic ingredients or cosmological assumptions, arising instead directly from the intrinsic geometry of shape space and the scale-invariant character of gravitational interactions. This perspective opens the possibility that a relational, scale-invariant reformulation of Newtonian dynamics could provide new conceptual bridges between gravitational dynamics, cosmology and the emergence of structure in the universe.

{\it Scale Invariance}:
As mentioned above, scale invariance has played only a marginal and largely accidental role in the historical development of physics, appearing in Newtonian gravity primarily through total-collision solutions, first identified by Sundman in the three-body problem \cite{sundman1907} and later extended to the $N$-body case \cite{block1909_chocs}. At the same time, Poincaré clearly recognized the empirical indistinguishability of global rescalings \cite{poincare1952_science}, emphasizing that only relations---ratios and angles---are physically observable. This observation anticipates a relational view in which scale has no intrinsic meaning and physical laws are invariant under global dilatations.

Scale invariance is in fact recognized in dynamics, not as a fundamental principle but in situations in which the potential is homogeneous of some degree. Then, as is well known, the equations of motion are invariant under the anisotropic rescaling: 
\begin{equation}
	\mathbf{r}_a \rightarrow \alpha \mathbf{r}_a, \qquad t \rightarrow \alpha^{3/2} t
\end{equation}
with $\alpha > 0$. This means that if $\{\mathbf{r}_a(t)\}$ is a solution, then the rescaled trajectories $\{\alpha\mathbf{r}_a(\alpha^{3/2}t)\}$ also solve the equations of motion. This symmetry is known as dynamical similarity and it has a familiar consequence: Kepler's third law. 

While this example does show that scale invariance is important under certain conditions, it and other such examples (see, e.g., \cite{sloan2018_ds}), have not yet prompted the proposal that scale invariance could serve as a first principle in physics and that primacy should be given to relations as opposed to absolute quantities such as size. In fact, to assign any meaning to size we always have to rely on a reference frame. In a universe without any fixed external reference, scale cannot be defined in any significant way. All that remain are dimensionless ratios between quantities intrinsic to the system itself. 

What is perhaps most intriguing is that scale invariance appears to offer the basis for a first-principles theory of creation, here understood as its etymology from the Proto-Indo-European root \textit{ker}, meaning ``to grow'', suggests it should. The core of the $N$-body problem is hidden by Newton's formulation of its equations in absolute space (and with it, an implicit, seldom recognized, absolute scale) and time \cite{barbour2025_talk}. When that reductionist framework is removed and no longer constrains behavior, one sees that even the slightest non-uniformities give rise to filaments, clusters, rings and other structures that keep on growing because the scale-invariant interactions drive the system to regions of shape space with ever greater variety. The evolution of the universe is a perpetual act of creation: not \textit{ex nihilo} but as the ceaseless growth and unfolding of shapes.

{\it Shapes}:
To clarify what is meant by shape, consider the simplest physical model: point particles and their spatial separations. In this model, a universe consisting of three particles defines a triangle. From the distances between these particles, one can form scale-invariant ratios that uniquely determine the internal angles of the triangle, thereby specifying its intrinsic shape. This idea naturally extends to any number of particles and underlies our relational perspective of the universe.

Observationally, cosmological claims are always grounded in ratios: the inference of cosmic expansion relies on the growth of the ratio between intergalactic separations and galactic diameters, while redshift itself is a ratio of wavelengths. These considerations suggest that what we observe is not expansion in an absolute sense, but rather a change in the shape of the universe. Therefore, we argue that only ratios can be accessed by observers within the universe.

{\it Variety}:
Consider a distribution of $N$ particles\footnote{We plan to extend the framework to an infinite number of particles.} with masses expressed as fractions of a total mass, $M$, and Cartesian coordinates, $\mathbf{r}_i$, in space. We introduce the variety function through a question: what simple scale-invariant quantity that takes into account all particles on an equal footing can be used to characterize the extent to which they are uniformly distributed or clustered?

Given our ontology of point particles and their spatial separations, the simplest way to define a scale-invariant function is as the ratio of two quantities with the dimension of length. We therefore choose as candidates the root-mean-square length, $\ell_{rms}$, 
\begin{equation}\label{eq2}
	\ell_{rms} = \frac{1}{M}\sqrt{\sum_{i<j}m_im_jr^2_{ij}},
\end{equation}
and the mean-harmonic length, $\ell_{mhl}$, 
\begin{equation}\label{eq3}
	\ell^{-1}_{mhl} = \frac{1}{M^2}\sum_{i<j} \frac{m_im_j}{r_{ij}},
\end{equation}
where $r_{ij} = ||\mathbf{r}_i - \mathbf{r}_j||$. 

The variety\footnote{Also referred to as complexity, as in \cite{barbour2014_arrow, barbour2024_complexity}.} is then defined as the ratio:
\begin{equation}
	V := \frac{\ell_{rms}}{\ell_{mhl}}.
\end{equation}
Beyond their simplicity, both the root-mean-square length and the mean-harmonic length are special. First, $\ell^2_{rms}$ is proportional to the center-of-mass moment of inertia, $I_{cm}$, which is a measure of the size of the system. For its part, $\ell_{mhl}$ is proportional to the negative inverse of the Newtonian potential, $V_{New}$, with $G = 1$. Thus, our candidate numbers are precisely the two that characterize Newtonian universal gravitation. Therefore, the variety also reads\footnote{If we normalize the total mass to unity, then equations \ref{eq2} and \ref{eq3}, as well as the expressions for the variety, simplify considerably.}:
\begin{equation}
	V = - \frac{1}{M^{5/2}}\sqrt{I_{cm}}V_{New}. 
\end{equation}

Our $V$, which in $N$-body theory is called the shape potential or the normalized Newtonian potential, is a sensitive measure of clustering because close approach of a few particles hardly changes $\ell_{rms}$ but greatly reduces $\ell_{mhl}$, increasing $V$ accordingly. It is striking that the purely mathematical measure of structural variety gives rise to and controls gravity.

Moreover, $V$ determines the intrinsic size of the particle distribution it characterizes, because $\ell_{rms}$ represents the average of the large separations in a distribution of points, while $\ell_{mhl}$ represents the average of the small separations. Therefore their ratio, which is $V$, yields a scale intrinsic to the system itself. No extrinsic ruler, such as $\ell_{rms}$, is required. 

It might be argued that $V$ is obtained through an artificial marriage of two physically distinct quantities: the square root of the center-of-mass moment of inertia and the Newtonian potential. However, given our suggestion that any dynamical system should be characterized in dimensionless terms, it is $V$ that is physical and splitting it into $\sqrt{I_{cm}}$ and $V_{New}$ is artificial.

{\it Central Configurations}:
In Newtonian gravity, central configurations (CCs) play a fundamental role in the study of the $N$-body problem. We say that a configuration of $N$ bodies is central if the acceleration vector of each body, resulting from mutual gravitational interactions, is proportional to its position vector relative to the system’s center of mass. Formally, central configurations are defined as sets of $N$ bodies with positions $\mathbf{r}_i \in \mathbb{R}^3$ and masses $m_i$, where $i=1,...,N$, such that, for all $i$,
\begin{equation}
	\sum_{j\neq i} G \frac{m_j (\mathbf{r}_j - \mathbf{r}_i)}{||\mathbf{r}_j - \mathbf{r}_i||^3} = -\lambda (\mathbf{r}_i-\mathbf{r}_{cm}),
\end{equation}
where $r_{cm}$ denotes the center of mass, $G$ is the gravitational constant and $\lambda$ a positive constant. 

Central configurations give rise to both homothetic and relative equilibria solutions of the $N$-body problem. The Lagrange points in the three-body problem correspond to the positions associated with such configurations. For more details about CCs see \cite{saari_cc}.

For us, the most important property of central configurations is that they are the critical points of the variety---minima and saddles, whose number grows faster than factorially with $N$. CCs are easily found numerically. A well-established result is that CCs at or very close to the absolute minimum of $V$ have a very uniform distribution of particles within a sphere with an abrupt boundary. For instance, Battye et al. \cite{battye2002_central} performed numerical calculations for the equal-mass case, showing that the absolute minimum of $V$ for $N\approx 10^3-10^4$ corresponds to extraordinarily uniform shapes.

At critical points, $V$, as a product of potentials, generates both attractive gravitational forces and repulsive Hooke forces with strengths precisely balanced to maintain the particles in equilibrium. However, away from critical points, this balance between the attractive Newtonian contribution and the repulsive Hooke-type contribution is broken, and one force becomes dominant. When the repulsive force dominates, a net repulsive effect emerges. We suggest this is formally analogous to Einstein’s cosmological constant, which was introduced \textit{ad hoc} to obtain a static universe. In our approach, by contrast, it arises directly from scale invariance. 

\begin{figure*}[ht!]
	\centering
	\includegraphics[width=\columnwidth]{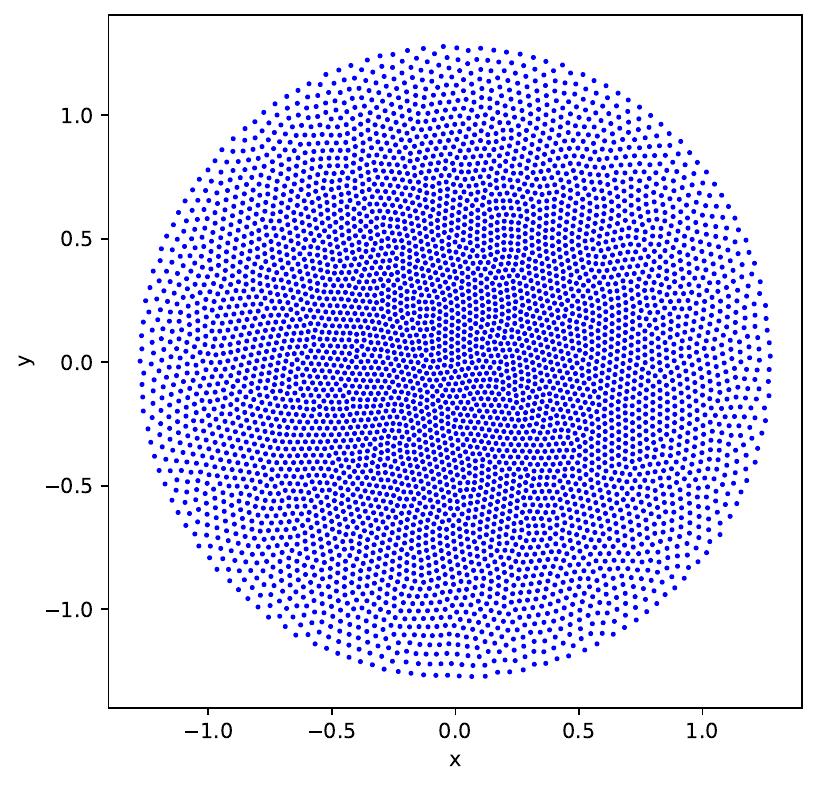}
	\vspace{0.3cm}
	\includegraphics[width=\columnwidth]{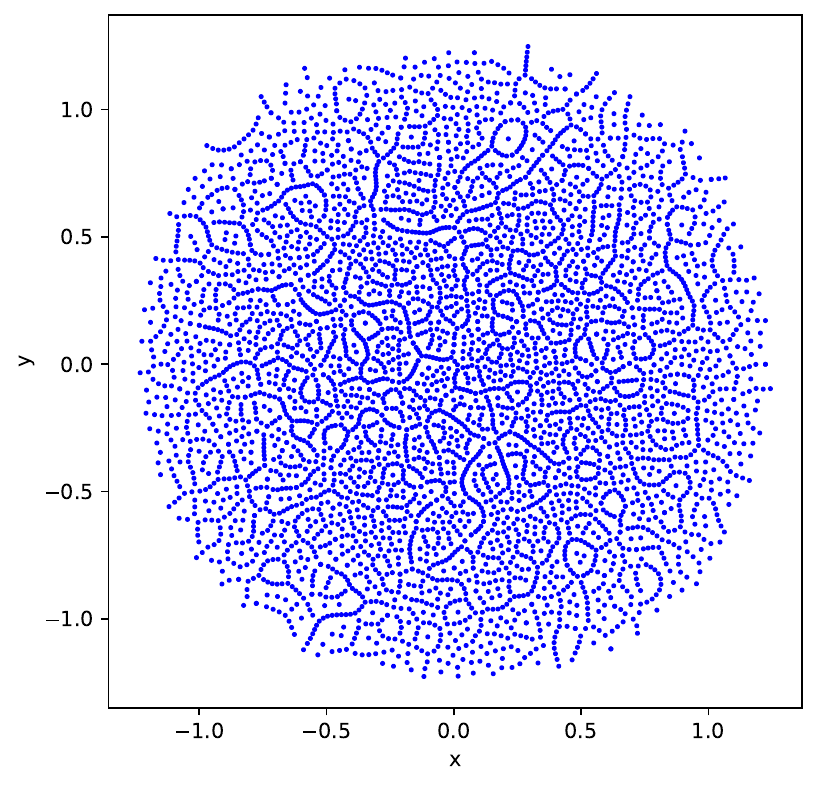}
	\caption{Two-dimensional central configurations of 5000 equal-mass particles with varieties very close to the absolute minimum of $V$ (left) and about 1.5\% above it (right).}
	\label{1}
\end{figure*}

Numerical studies have essentially been focused on locating CCs at, or as close as possible to, the absolute minimum of $V$. The interest of one of us [JB] in shapes and the fact that, under the time-reversal symmetry of Newton's laws, total-collision solutions can be seen as Newtonian big bangs, led Manuel Izquierdo under the supervision of Alain Albouy at the Paris Observatory to look for central configurations slightly above the absolute minimum of the variety. This culminated in the discovery of the planar CCs presented in \cite{izquierdo2021_filaments, barbour2024_complexity}. One of us [MIRL] reproduced those results for a larger system of particles (Fig.~\ref{1}). The resulting patterns bear a striking resemblance to the cosmic web. 

This behavior is a novelty. Filamentary structures emerge already when $V$ is little more than 1\% above its absolute minimum. We also observe closed loops comprising anywhere from zero to ten or more particles, a multitude of small-scale structures and particles roughly uniformly distributed throughout the larger filamentary network. The increase in particle separations from the center toward the periphery in Fig.~\ref{1} is due to the calculations having been made in two dimensions.

\begin{figure*}[ht!]
	\centering
	\includegraphics[width=\columnwidth]{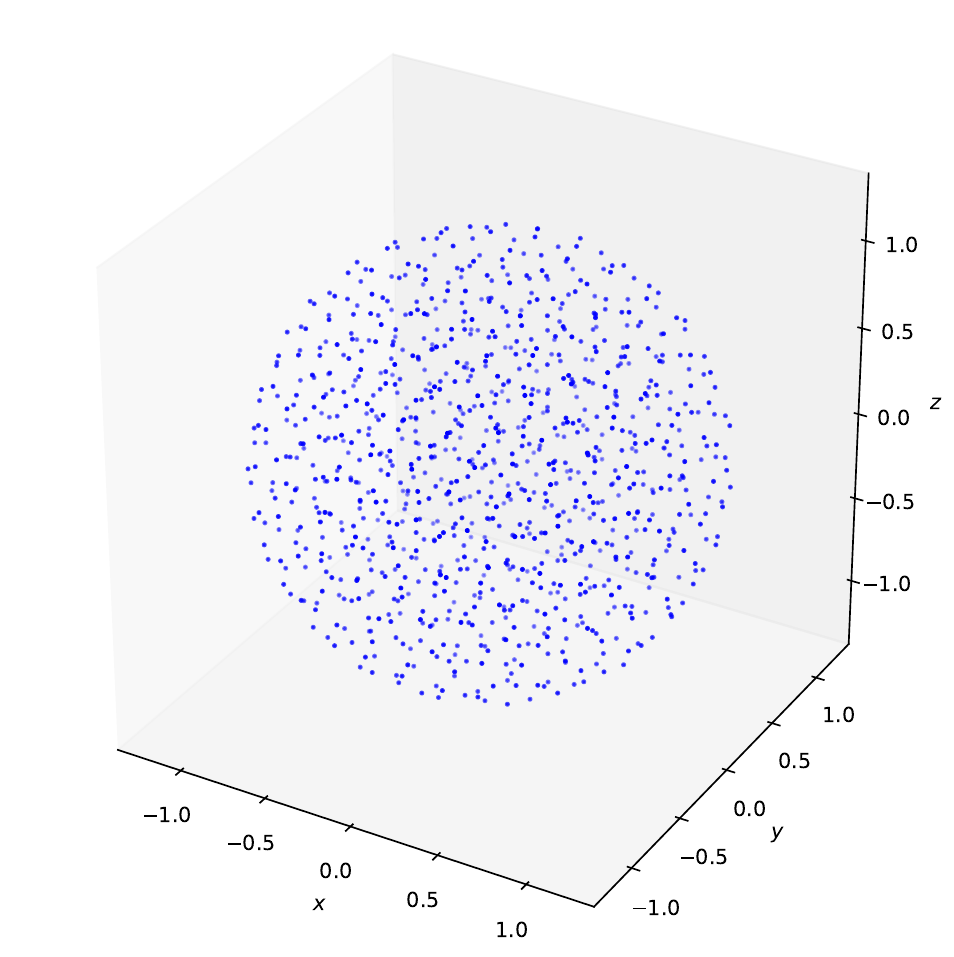}
	\vspace{0.3cm}
	\includegraphics[width=\columnwidth]{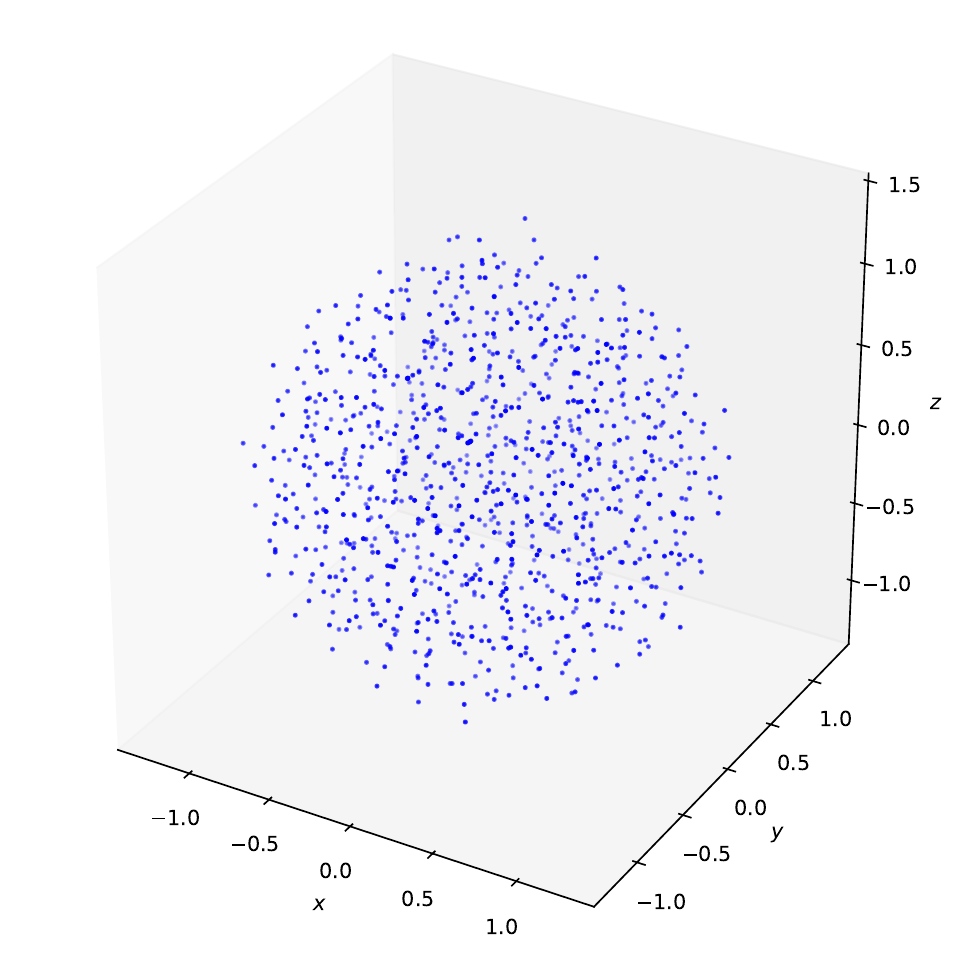}
	\caption{Three-dimensional central configurations of 1000 equal-mass particles with varieties very close to the absolute minimum of $V$ (left) and about 1.5\% above it (right).}
	\label{2}
\end{figure*}

\begin{figure*}[ht!]
    \centering
    \includegraphics[scale=0.483]{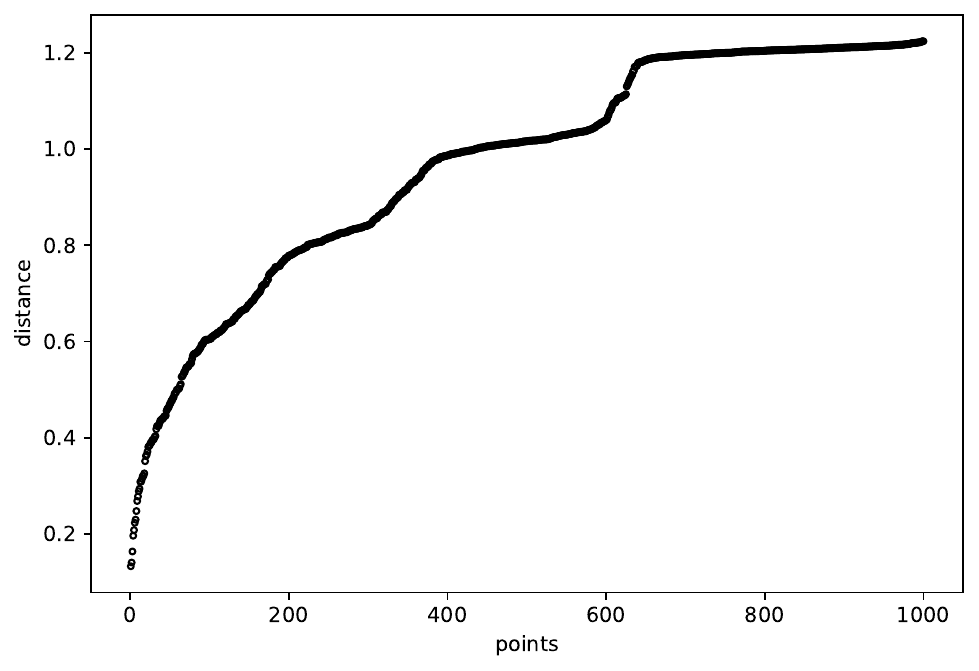}
    \vspace{0.3cm}
    \includegraphics[scale=0.483]{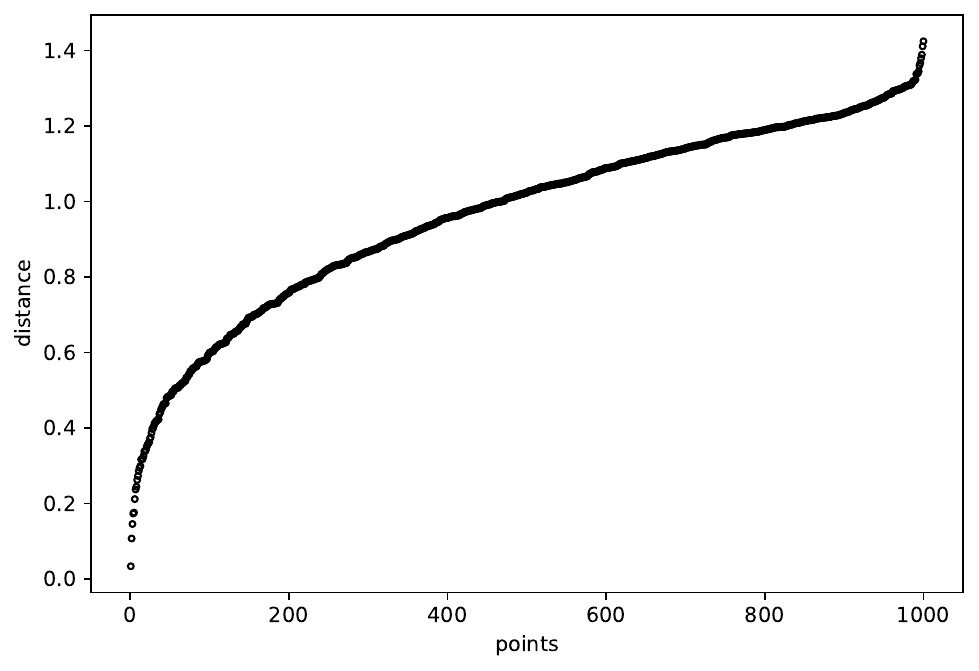}
	\caption{The distance of each point from the center of mass for the central configurations near the absolute minimum of $V$ (left) and $1.5\%$ above it (right).}
	\label{3}
\end{figure*}

\begin{figure*}[ht!]
    \centering
    \includegraphics[scale=0.483]{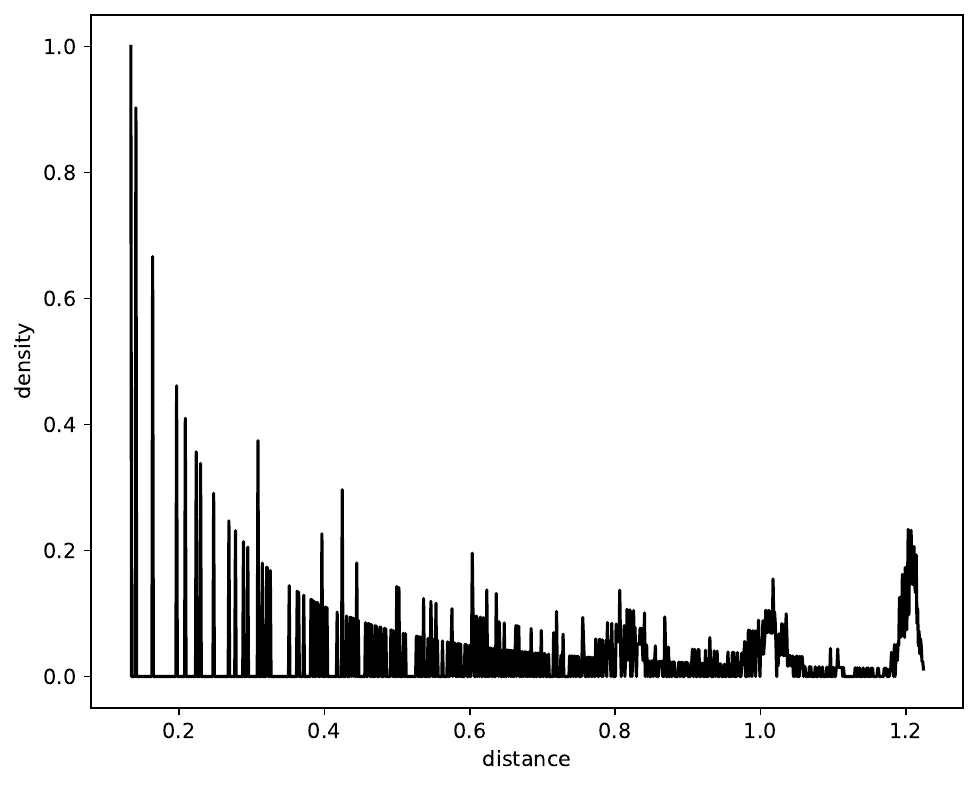}
    \vspace{0.3cm}
    \includegraphics[scale=0.483]{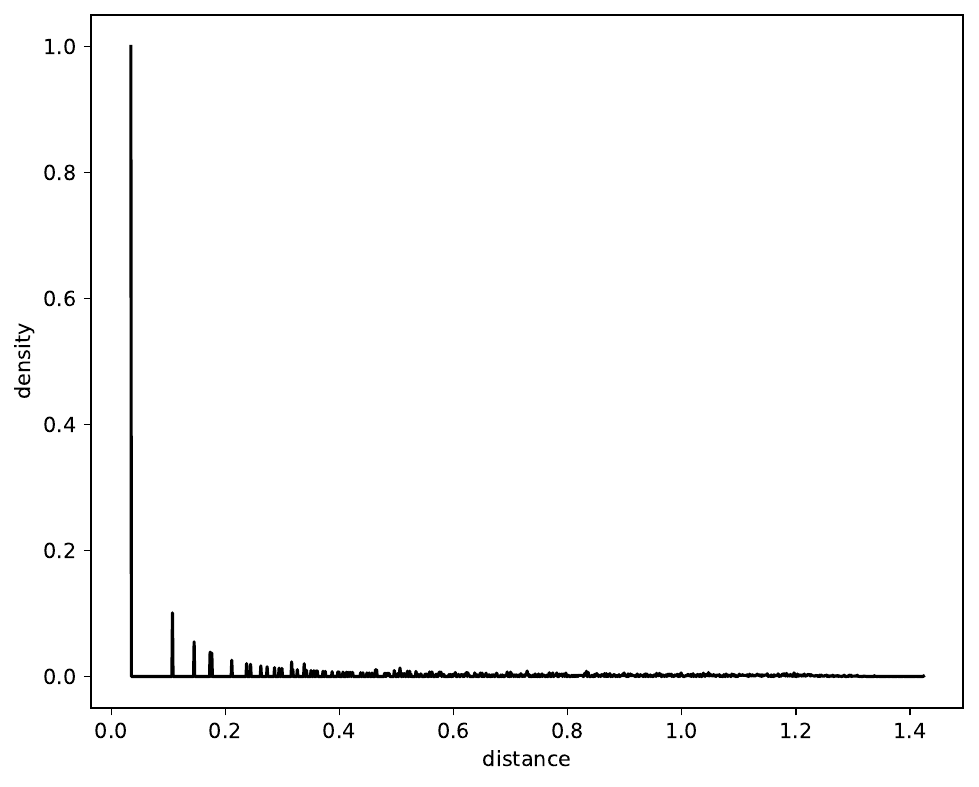}
	\caption{The density as a function of radial distance for the central configurations near the absolute minimum of $V$ (left) and $1.5\%$ above it (right).}
	\label{4}
\end{figure*}

\begin{figure*}[ht!]
    \centering
    \includegraphics[scale=0.483]{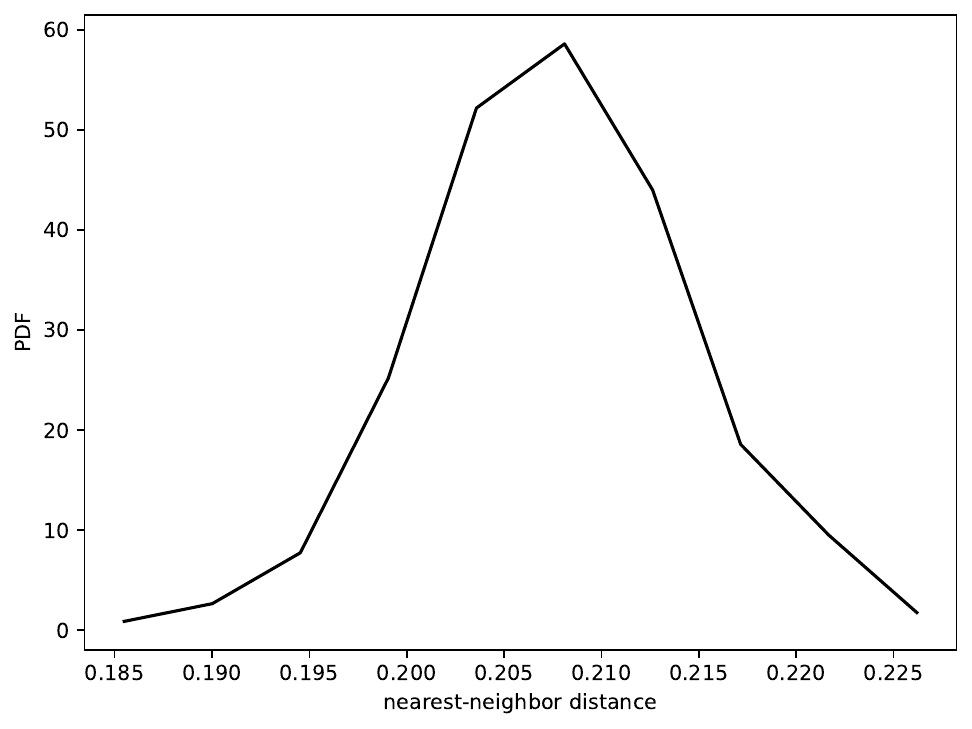}
    \vspace{0.3cm}
    \includegraphics[scale=0.483]{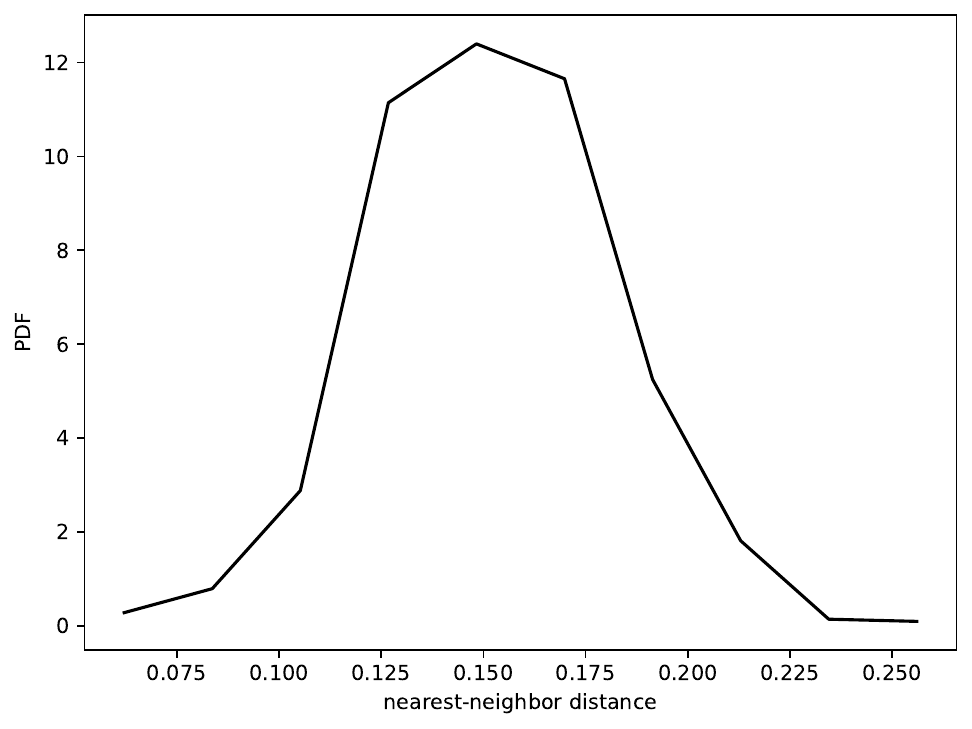}
	\caption{Nearest-neighbor distribution for the central configurations near the absolute minimum of $V$ (left) and $1.5\%$ above it (right).}
	\label{5}
\end{figure*}

Recent numerical studies in three dimensions performed by one of us [MIRL] confirm a similar behavior. Although more challenging to visualize, comparison between the two spatial CCs in Fig.~\ref{2} clearly shows a departure from uniformity from the CC on the left to the one on the right. They were obtained using a similar numerical approach to that in \cite{izquierdo2021_filaments}. We will call Configuration 1 the configuration near the absolute minimum of $V$ and Configuration 2 the one above it. Figures \ref{3}, \ref{4} and \ref{5} exhibit, respectively, the distance of each particle to the center of mass of the system, the density as a function of the distance and the nearest-neighbor distribution for both configurations in Fig.~\ref{2}. In three dimensions it becomes more difficult to find the absolute minimum of the variety. Nevertheless, our results for the configuration with lowest variety are in good agreement with those reported in \cite{battye2002_central}.

The radial distance, depicted in Fig.~\ref{3}, provides a first global characterization of the spatial organization of the two configurations. For Configuration 1, the plot reveals the presence of concentric layers, or shells, of densely packed particles. This structure is consistent with a relatively uniform yet radially ordered arrangement, in which particles preferentially occupy specific distances from the center of mass. This behavior is absent in Configuration 2, where the radial distance distribution is smooth, indicating that particles do not accumulate at well-defined radial positions. In \cite{battye2002_central}, the authors have found a configuration closer to the absolute minimum, in which the formation of concentric shells is even more pronounced.

The analysis of the radial density profiles, shown in Fig.~\ref{4}, is fully consistent with the previous results and further highlights the distinction between the two regimes. In both configurations, the density decreases with increasing distance from the center of mass. However, in Configuration 2 the density exhibits a drastic exponential decay, indicating strong spatial localization and the absence of long-range radial structure. By contrast, Configuration 1 displays a slower decay modulated by periodic peaks. These peaks coincide with the layers identified in the radial distance plot depicted in Fig.~\ref{3}, reinforcing the interpretation of concentric shell formation and a more radially ordered configuration.

Finally, the probability density functions (PDFs) of the nearest-neighbor distances depicted in Fig.~\ref{5} reveal clear differences in the local structure of the two configurations. In Configuration 1, nearest-neighbor distances are narrowly distributed around a well-defined characteristic scale, meaning a small variation in the distances. The distribution is mildly asymmetric, with a longer tail toward larger distances, indicating a predominantly uniform local arrangement. In contrast, Configuration 2 exhibits a broader distribution shifted toward smaller distances. This shift toward shorter nearest-neighbor distances is consistent with filament formation, in which particles aggregate along elongated structures, leading to reduced typical separations at the local level.

All these results suggest a novel and rich behavior also in three dimensions, where the intrinsic structure of shape space drives the system toward regions of higher variety---equivalently, lower potential---leading to the spontaneous formation of filamentary structures reminiscent of the large-scale cosmic web.

\textcolor{white}{Este texto foi adicionado para a próxima secção começar numa página nova.}

{\it Gravitational Arrow of Time}:
It is commonly assumed that any arrow of time must be imposed through special initial conditions on otherwise time-reversal invariant laws. This view is challenged by the Newtonian $N$-body problem with vanishing total energy and angular momentum, which provides a dynamically closed, time-symmetric model of a universe. As shown by Barbour, Koslowski and Mercati \cite{barbour2014_arrow}, almost all solutions of this system possess a unique point, denoted as the \emph{Janus point}, at which $I_{cm}$ attains a global minimum. This point divides each solution into two time-reversed halves. Away from it, a dimensionless measure of shape inhomogeneity grows irreversibly, despite the exact time-reversal symmetry of the equations of motion. No special initial conditions are required: the arrow of time emerges generically from the structure of the dynamics itself.

In the scale-invariant formulation adopted here, this measure is precisely the variety $V$, the normalized Newtonian potential and a function defined intrinsically on shape space. The absolute minimum of $V$ corresponds to extraordinarily uniform configurations occupying a negligibly small region of shape space. Generic solutions therefore pass briefly through such low-variety states near the Janus point and subsequently evolve toward regions of ever higher variety. As $V$ increases, dynamically stable subsystems, such as, clusters, binaries and filamentary structures, are created and act as records, storing information about past interactions. Any internal observer must inhabit one branch of the solution and will infer a unique past and future direction from these records, identifying the direction of increasing $V$ as the future. The gravitational arrow of time thus arises as a relational, scale-invariant phenomenon: a direct consequence of the geometry of shape space and the unbounded growth of variety, rather than of entropy maximization, cosmological expansion or imposed temporal asymmetry.

{\it Conclusion}:
In this work we have treated scale invariance not as an incidental symmetry but as a guiding principle for gravitational dynamics. By focusing on relational, scale-free quantities, we showed that the Newtonian $N$-body problem admits a natural formulation on shape space, where physical evolution is described in terms of intrinsic, dimensionless variables rather than absolute size or external reference structures.

Central to this framework is the variety $V$, a dimensionless quantity that coincides with the normalized Newtonian potential of $N$-body theory and provides a meaningful measure of structure and clustering in closed gravitational systems. Central configurations arise as the critical points of $V$, with the absolute minimum corresponding to maximally uniform states. Numerical results in two and three dimensions show that small departures from this minimum generically lead to the spontaneous emergence of filamentary structures, without additional forces or fine-tuned initial conditions, as a direct consequence of the geometry of shape space and the scale-invariant nature of gravity.

Finally, the scale-invariant formulation developed here yields a gravitational arrow of time without special initial conditions: generic solutions of the dynamically closed Newtonian $N$-body problem pass through a unique point of minimal variety and then evolve toward regions of higher $V$. The secular growth of variety simultaneously generates structure and records, making temporal directionality and structure formation inseparable consequences of scale-invariant dynamics on shape space.

The framework developed here opens several promising directions for future research. On the mathematical side, a more systematic exploration of the landscape of $V$, particularly in three dimensions and for unequal masses, could shed light on the organization of shape space and the statistics of emergent structures. 

From a cosmological perspective, the results motivate a reassessment of standard interpretive frameworks, suggesting that expansion, structure formation and possibly dark energy may admit relational reinterpretations within a scale-invariant setting, even though precise connections with relativistic cosmology remain an open challenge.

More broadly, this work supports the view that relationalism and scale invariance provide not merely philosophical alternatives to absolutist formulations, but concrete tools for uncovering new physical behavior. Whether this perspective can be extended beyond Newtonian gravity to quantum theory or general relativity remains an open question but the results presented here offer a compelling starting point.\\

{\it Acknowledgements}:  
MIRL acknowledges the Instituto de Astrofísica e Ciências do Espaço (IA) for providing the computer cluster used in this work, as well as valuable feedback from J. E. C. Carmelo.
JB thanks Tim Koslowski for many helpful discussions.
FSNL acknowledges support from the Fundação para a Ciência e a Tecnologia (FCT) through a Scientific Employment Stimulus contract (reference CEECINST/00032/2018), as well as funding from the research grants UID/04434/2025 and PTDC/FIS-AST/0054/20.

\end{document}